\magnification=\magstep1
\baselineskip=13pt
\parskip = 3pt

\font\title = cmbx10 scaled 1440        

\centerline{{\title On the absence of a measurement problem}}
\vskip 5pt
\centerline{{\title in quantum computer science}}
\vskip 10pt
\centerline{N. David Mermin}
\centerline{Laboratory of Atomic and Solid State Physics}
\centerline{Cornell University, Ithaca, NY 14853-2501}
\vskip 10pt

{\narrower \narrower

\noindent I comment on GianCarlo Ghirardi's criticism of my claim that 
quantum computation has no measurement problem.

}
\vskip 10pt

In view of GianCarlo Ghirardi's remarks (arXiv:0806.0647)
about my {\it In Praise of Measurement} (arXiv:quant-ph/0612216), I
should clarify what I meant to say in that paper and in the conference
lecture on which it was based.  I wanted to make two points about
measurement in quantum computer science.  The first was that the simple and
crucial role of measurement gates deserved greater emphasis:

{\narrower

\noindent (a) The only measurement gates ever needed are multiple copies of a
single one, that acts on a single Qbit.  All other processes called
``measurement'' can be built out of 1-Qbit measurement gates and
unitary gates. The question of what qualifies as a measurement has a
precise answer.

\noindent (b) Applying and reading measurement gates is an
indispensable part of any computation.  If gates were restricted to
unitary gates, then quantum computation would be useless, since no
user could acquire any numerical information.

\noindent (c) In addition to creating the output of the computation,
measurement gates are the most conceptually straightforward and
pedagogically economical way to get the computation underway, by, for
example, enabling the user to assign to each Qbit the initial state
$|0\rangle$.

}

My second point was that if quantum mechanics were {\it only\/} a branch of
computer science that applied {\it only\/} to the operation of quantum
computers, then there would be no measurement problem.  This
conclusion grew out of my experience teaching quantum mechanics to
computer scientists for 6 years, during which no ``measurement
problem'' ever arose in the course of innumerable discussions with
students trying to learn the subject without any prior knowledge of
quantum mechanics, and during which I, as teacher, never felt
dissatisfied with what I was able to tell the students, as I do at
various points when teaching a physics course in conventional quantum
physics. For more on this see Part 3 of {\it Copenhagen Computation:
How I Learned to Stop Worrying and Love Bohr\/}
(arXiv:quant-ph/0305088).

Ghirardi objects to my second point, making me realize that I should
have been more emphatic that the claim in my article that there is no
measurement problem is limited to quantum computer science.  My
abstract says ``I argue that within the field of quantum computer
science the concept of measurement is $\ldots$ unproblematic
$\ldots\,$.''  And in the paper I say my subject is ``the role of
measurement in quantum computation.''  And toward the end I say that
measurement ``is [not] problematic in quantum computer science.''
Only in the last five paragraphs do I raise the question of whether
``this generalizes beyond quantum computation'' and make some
tentative suggestions about how it might.

I should, however, have stated explicitly that I was not claiming that
quantum computation provided a solution to the broader measurement
problem, but only that it described a conceptual microcosm within
which there was nothing problematic about measurement. I did not
intend to imply that this ``weakened or even cancelled'' John Bell's
criticisms of quantum mechanics as a satisfactory picture of the
entire physical world.  On the contrary, at the beginning I say ``I'm
not sure Bell would have found any of the remarks that follow
compelling, or even suggestive.''  To underline my reservations about
how Bell might have reacted I voice the hope that Anton Zeilinger (the
honoree at the conference) might ``take a more sympathetic view of it
than Bell might have done."  After that, aside from asserting that Bell
would have enjoyed the quantum information revolution, I don't
speculate on how he might have responded, mentioning him only when I
note how some of his complaints about quantum foundations look in the
restricted setting of quantum computer science.  I brought
Bell into the story because the talk was at a conference of
people, most of whom were more involved in aspects of quantum
information, than they were in quantum foundations.  I could not
assume that all (or even most) of them had heard of a ``measurement
problem'', and could imagine no more effective way of introducing them
to the issues than to refer them to and quote from Bell's brilliant
and delightful essay, {\it Against Measurement.}

On the matter of Bell's use of ``exact", I was indeed thinking only of
his definition of ``exact" in {\it Against Measurement\/}: ``fully
formulated in mathematical terms, with nothing left to the discretion
of the theoretical physicist.''  I didn't know (or had forgotten)
that, as Ghirardi points out, elsewhere Bell uses the term to mean a
theory that ``neither needs nor is embarrassed by an observer."  But
had I known this, I would have added a remark to the effect that the
``user", though not always explicit in computer science, can hardly be
viewed as an embarrassment to the field.  (I know an information
scientist at Cornell whose specialty is ``The Computer-Human
Interface".  As far as I know, this does not embarrass her or her
colleagues.)  I do talk about the ``user", particularly in joking
about God and Einstein's mouse, but without realizing that I was
touching on another aspect of Bell's notion of ``exactness".

Having said this, I must acknowledge that Ghirardi correctly infers
that there are matters on which we disagree.  My own current view
is along the lines of the Bohr quotation I give at the beginning of
{\it In Praise of Measurement}: ``In our description of nature the
purpose is not to disclose the real essence of the phenomena but only
to track down, so far as it is possible, relations between the
manifold aspects of our experience.'' I take this to mean that quantum
states are not entities with an objective existence, or inherent
properties of the systems with which they are associated, but
mathematical constructs that enable us to relate some of our
experiences (e.g. what we read on the display of the measurement gates
that prepare the initial state) to others (e.g. what we read when we
look at the display of the final measurement gates). When we have new
experiences we update our state assignments accordingly, and this is
indeed a nonlinear alteration.  But what is altered is an abstraction
that provides a remarkably effective starting point for subsequent
calculation; it is not a real essence of the phenomena.

Admittedly, for this to be a coherent position, one must take some of
the manifold aspects of our experience to lie beyond the domain that
physics describes.  The concepts that physics has developed are
incapable of capturing the taste --- the indescribable flavor --- of
individual conscious experience. Physics has the goal of accounting
for correlations among different aspects of that (irreducible)
experience (which is all we could possibly need from it).  Quantum
computer science illustrates this strikingly and unproblematically
through a user, whose personal experience makes possible the initial
input and acknowledges the final output.

If there is a tricky conceptual issue within quantum computer science
it is not a measurement problem, but a unitary-evolution problem.
Measurements are a conceptually straightforward part of how you
initiate and conclude the computation.  But the state that evolves
unitarily after the initial and before the final measurements is more
subtle.  Students (and many contemporary physicists, including, I
believe, Ghirardi) want to regard the state as a real essence of the
phenomena --- an objective property of the Qbits (as we are used to
regarding the state of classical Cbits) --- even though there is no
way to determine what the state is, given only the Qbits, and even
though that view leads unambiguously to unmediated (spooky) action at a
distance.  In teaching quantum computation to computer science
students (and in my book {\it Quantum Computer Science: An
Introduction\/}) I stress that the state and the unitary
transformations it is subject to are mathematical abstractions that
enable one to compute from a knowledge of the readings of the initial
measurement gates and the components of the circuit that follows, the
probabilities of the readings of the final measurement gates.
Mathematical abstractions do not require stochastic ``hits''
originating in unknown physical processes (or interactions
with gravitons) to be reset; they are reset by us, when we acquire
more information and want to calculate what we can expect to
experience next.

So I do maintain that there is no measurement problem within quantum
computer science.  And I very much wish that John Bell were still
among us to object.

\bye